\documentclass[prb]{revtex4}
\usepackage{epsfig}
\usepackage{amsmath}
\begin{document}

\title{Phase rigidity breaking in open Aharonov-Bohm ring coupled to a cantilever}
\author{F. Romeo, R. Citro and M. Marinaro}
\affiliation{Dipartimento di Fisica ``E. R. Caianiello",
Universit{\`a} degli Studi di Salerno, and Unit{\`a} C.N.I.S.M.,
Via S. Allende, I-84081 Baronissi (SA), Italy}

\date{\today}

\begin{abstract}
The conductance and the transmittance phase shifts of a
two-terminal  Aharonov-Bohm (AB) ring are analyzed in the presence
of mechanical displacements due to coupling to an external
cantilever. We show that phase rigidity is broken, even in the
linear response regime, by means of inelastic scattering due to
phonons. Our device provides a way of observing continuous
variation of the transmission phase through a two-terminal
nano-electro-mechanical system (NEMS). We also propose
measurements of phase shifts as a way to determine the strength of
the electron-phonon coupling in NEMS.
\end{abstract}



\maketitle

Nanoelectromechanical systems (NEMS) have been a subject of
extensive research in recent years. The possibility of combining
electrical and mechanical degrees of freedom on the nanoscale may
give rise to technological advantages as well as manifestations of
fundamental physical phenomena. From a technological point of view
the interest is largely due to the many applications that may be
realized using NEMS.\cite{cleland_review_NEM} It follows
ubiquitously from the concrete possibility of downsizing devices
from micromechanical into nanoelectromechanical systems. Among the
many NEMS phenomena of considerable physical interest, we focus in
this paper on the effect of quantum-coherent displacements in the
presence of a Aharonov-Bohm effect in a one-dimensional ring
symmetrically connected to two external leads. On the nanoscale
level the mechanical forces controlling the structure of the
system are of the same order of magnitude as the capacitive
electrostatic forces governed by charge distributions. This
circumstance is of the utmost importance when analyzing electron
tunneling. For example, some predictions on the role of
electron-phonon interactions on the electron tunneling have
regarded the modifications of the peak-to-valley current ratio due
to inelastic processes with important consequences in device
applications\cite{goldman_dephasing,dephasing_cond}. Another
important issue in NEMS physics is related to the way mechanical
displacements perturb phase coherent charge transport through
closed loops as a consequence of the shifting of electron
trajectories. For example, important information come from the
Aharonov-Bohm (AB) oscillations in mesoscopic
rings\cite{yacoby_AB,buks_AB,stern_AB,yeyati_AB,feng_AB}. Here,
the loss of coherence, or dephasing, due to inelastic scattering
caused by phonons reveals in the suppression of $hc/e$
oscillations. Another important aspect unnoticed in previous
studies of NEMS is that inelastic scattering breaks
phase-rigidity, i.e. the linear conductance is not symmetric in
the AB phase. Here we show that the observed asymmetry can be
tuned continuously by changing the electron-phonon coupling,
demonstrating that the phase of the linear conductance in a
two-terminal AB interferometer is not rigid when tunneling is
assisted by phonons. The problem of phase rigidity and its
breaking in ring and ring-dot systems has been largely
investigated both
theoretically\cite{imry_review,jayannavar,yeyati_AB,phase_rig_1,phase_rig_2,phase_rig_3,sun99}
and
experimentally\cite{buks_exp,exp_kobayashi1,exp_kobayashi2,leturcq06exp}.
Here we address for the first time the problem in the framework of
nanoelectromechanical systems and propose a way to detect phase
rigidity breaking.

We consider a one-dimensional ring symmetrically coupled to two
leads and to a mechanical cantilever. A coupling between the
electrons travelling in the lower arm of the ring and the
cantilever, whose tip is suspended over the arm, can be set up by
developing a uniform electric field between the tip, the
cantilever and the lower arm. Electrons couple approximately
linearly to the cantilever position, thus leading to a coupling
between the flexural phonon modes of the cantilever and the local
density of the electrons on the ring arm. Furthermore, because the
coupling strength decays rapidly with increasing frequency, for
micron scale cantilevers only the fundamental flexural mode is
relevant. Therefore, at low enough temperature, the cantilever can
be treated as a single quantum mechanical oscillator. Of course
the dephasing behavior of the electrons due to the cantilever
depends on the relative magnitudes of the electron dwell time on
the lower arm of the ring and the cantilever period. A similar
device has been proposed by A. MacKinnon and A. D. Armour as a
which-path device for electrons [\onlinecite{armour03}].

\begin{figure}[htbp]
\centering
\includegraphics[scale=.55]{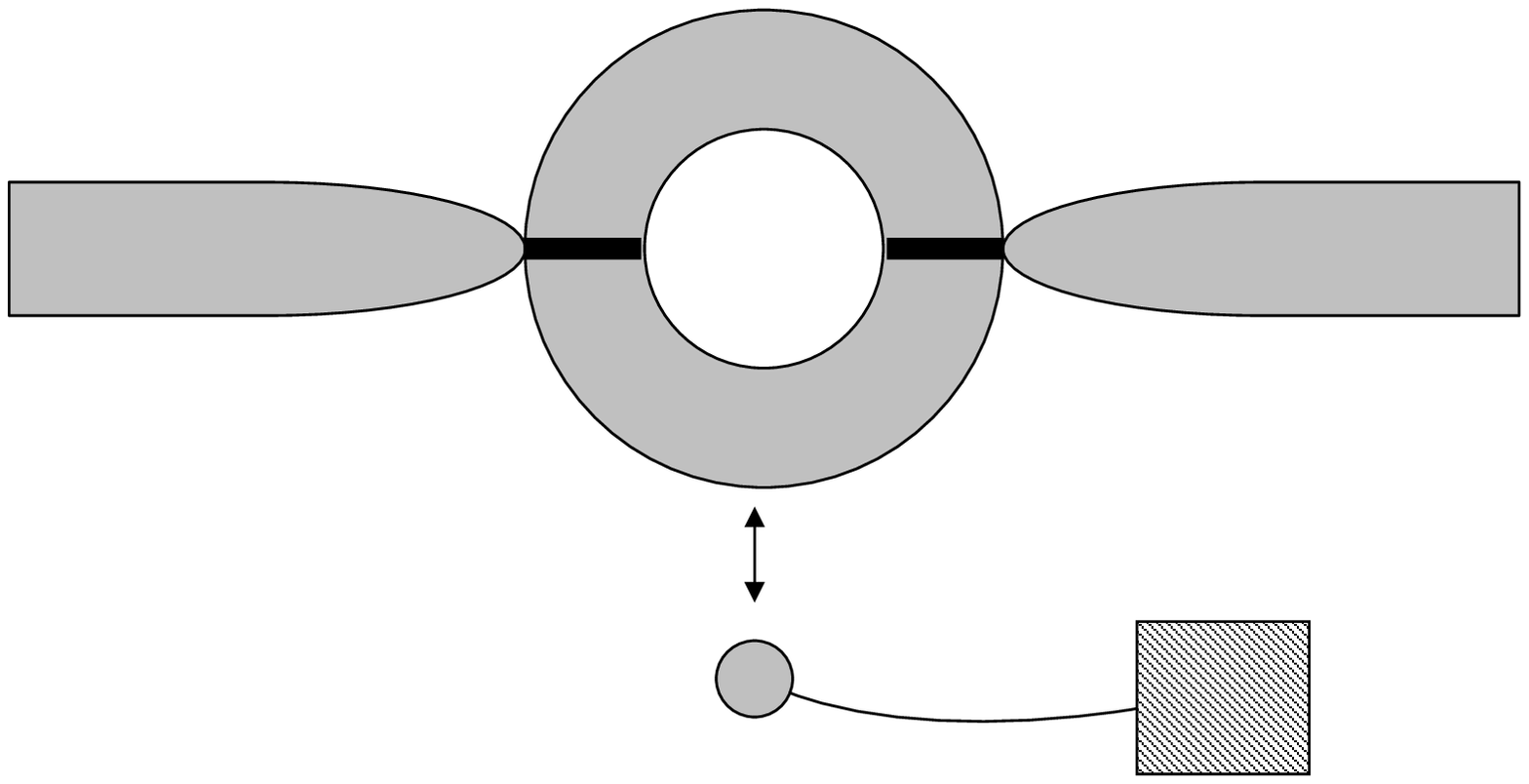}\\
\caption{Device: A mesoscopic ring symmetrically coupled to two
external leads and  a cantilever.} \label{fig:device}
\end{figure}

In the following we generalize the scattering matrix approach in a
way suitable for electron dwell times shorter or comparable to the
cantilever period. In this way both elastic and inelastic
contributions to the scattering process will appear.
The Hamiltonian of the system shown in Fig.\ref{fig:device} is
written as:
\begin{equation}
\mathcal{H}=\mathcal{H}_r+\mathcal{H}_c+\mathcal{H}_{int},
\end{equation}
where  $\mathcal{H}_r$ and $\mathcal{H}_c$ are the Hamiltonian of
the ring (in polar coordinates) and the cantilever, respectively:
\begin{eqnarray}
\mathcal{H}_r= \frac{\hbar^2}{2m_e R^2}\left( -i
\frac{\partial}{\partial
\varphi}-\frac{\Phi_{AB}}{\phi_0}\right)^2\\
\mathcal{H}_c=\frac{p^2_y}{2M_c}+\frac{M_c\omega^2 y^2}{2},
\end{eqnarray}
while $\mathcal{H}_{int}$ describes the interaction between the
lower arm of the ring ($0< \varphi< \pi$) and the cantilever
\begin{equation}
\mathcal{H}_{int}(x=R\varphi,y)=
  \begin{cases}
    \alpha y & \text{$0< \varphi< \pi$}, \\
    0 & \text{otherwise},
  \end{cases}
\end{equation}
where $\alpha$ represents the average of the electron density on
the lower arm of the ring times the uniform electric field. The
interaction between the electrons on the lower arm of the ring and
the cantilever is modelled as a linear coupling between the
displacement of the flexural mode and the average electron density
of the lower arm of the ring, since when electrons are not in the
ring there is no displacement of the cantilever (we assume that
all positions are measured from the equilibrium height of the
cantilever). The linear coupling is a valid approximation in the
limit in which the displacement of the cantilever from its
equilibrium height is small compared to the characteristic length
of the harmonic oscillator. The Hamiltonian of left and right lead
is given by the free particle Hamiltonian
$\mathcal{H}_{lead}=-\hbar^2\partial^2_{x^2}/(2m_{e})$. By
combining the harmonic potential of the cantilever with the
interaction term $\mathcal{H}_{int}$, the effective potential
along the lower arm can be rewritten as:
\begin{equation}
V_{eff}=\frac 1 2 M_c\omega^2
(y+\frac{\alpha}{M_c\omega^2})^2-\frac{\alpha^2}{2M_c\omega^2}.
\end{equation}
In the following we solve the scattering problem, generalized to
include inelastic scattering as in Ref.\onlinecite{bonca99}. For
our purposes, we set up a scattering problem by following the
method of quantum waveguide transport on
networks\cite{xia_waveguide,deo_94}. One main problem is the
boundary conditions at the intersection with the external leads.
In this case the Griffith boundary's
condition\cite{griffith_boundary} state that (i) the wave function
must be continuous and (ii) the current density must be conserved.
We assume that when an electron moves along the upper arm in the
clockwise direction from  $\varphi=0$, it acquires a phase
$\Phi_{AB}/2$ at the output intersection $\varphi=\pi$, whereas
the electron acquires a phase $-\Phi_{AB}/2$ in the
counterclockwise direction along the lower arm when moving from
$\varphi=0$ to $\varphi=\pi$. The wave-function for the right
(\textit{out}) /left (\textit{in}) lead and the upper
(\textit{up})/lower (\textit{low}) arm of the ring may be written
as:

\begin{eqnarray}
&&\psi_{in}(x,y)=\phi_l(y)e^{ikx}+\sum^\infty_{m=0}r_{ml}\phi_m(y)e^{-ik_m
x}\\
&&\psi_{out}(x,y)=\sum^\infty_{m=0}t_{ml}\phi_m(y)e^{ik_m
x}\\
&&\psi_{up}(x,y)=\sum^\infty_{m=0}\phi_m(y)(a_{ml} e^{ik^+_m x}+b_{ml}e^{-i \pi f}e^{-i k^-_m x})\\
&&\psi_{low}(x,y)=\sum^\infty_{m=0}\phi_m(y+y_0)(A_{ml}
e^{i\tilde{k}^-_m x}+B_{ml}e^{i \pi f}e^{-i \tilde{k}^+_m x}),
\end{eqnarray}
where $\phi_n(y)$ represents the $n-$phonon state, and we have
defined $f=\frac{\phi_{AB}}{\phi_0}$. The electron wavevectors,
determined by the energy conservation, are
$k_m=\sqrt{k^2+p^2(l-m)}$,
$\tilde{k}_m=\sqrt{k^2+p^2(l-m+\gamma^2)}$,
$p^2=2m_e\omega/\hbar$, $k^{\pm}_m=k_m \pm \pi f/L$,
$\tilde{k}^{\pm}_m=\tilde{k}_m \pm \pi f/L$ ($L=\pi R$ being the
half circumference). The phonon wave function in the lower arm is
shifted of the quantity $y_0=\frac{\alpha}{M_c\omega^2}$ and we
have defined the dimensionless parameter
$\gamma=y_0\sqrt{\frac{M_c\omega}{2\hbar}}$. The unknown
transmission and reflection coefficients are determined by solving
the following system of equations corresponding to the Griffith's
boundary conditions:
\begin{eqnarray}
\label{eq:linear_syst}
&&\psi_{in}(0,y)=\psi_{up}(0,y)=\psi_{down}(0,y)\nonumber \\
&&\psi_{out}(\pi R,y)=\psi_{up}(\pi R,y)=\psi_{down}(\pi R,y) \nonumber \\
&&\partial_x\psi_{up}(x=0,y)+\partial_x\psi_{down}(x=0,y)=\partial_x\psi_{in}(x=0,y)\nonumber\\
&&\partial_x\psi_{out}(x=\pi R,y)-\partial_x\psi_{up}(x=\pi
R,y)=\partial_x\psi_{down}(x=\pi R,y).
\end{eqnarray}
To eliminate the dependence on $y$, we project each equations on
the phonon state $\phi_s(y)$, the only non-trivial projection
being $\int^{\infty}_{-\infty}\phi_s(y)^{\ast}\phi_m(y+y_0)dy$.
Since $y_0$ is small quantity we may expand $\phi_m(y+y_0)$ as
$\phi_m(y+y_0)\approx
\phi_m(y)+\gamma(\sqrt{m}\phi_{m-1}(y)-\sqrt{m+1}\phi_{m+1}(y))$,
which allows immediately to get the following result:
\begin{equation}
\label{eq:approx}
\int^{\infty}_{-\infty}\phi_s(y)^{\ast}\phi_m(y+y_0)dy \approx
\delta_{s,m}+\gamma(\sqrt{m}\delta_{s,m-1}-\sqrt{m+1}\delta_{s,m+1}).
\end{equation}
Since $\gamma$ is a small quantity, only one-phonon processes are
relevant and the scattering problem can be solved with arbitrary
numerical accuracy.

\begin{figure}[htbp]
\centering
\includegraphics[scale=.55]{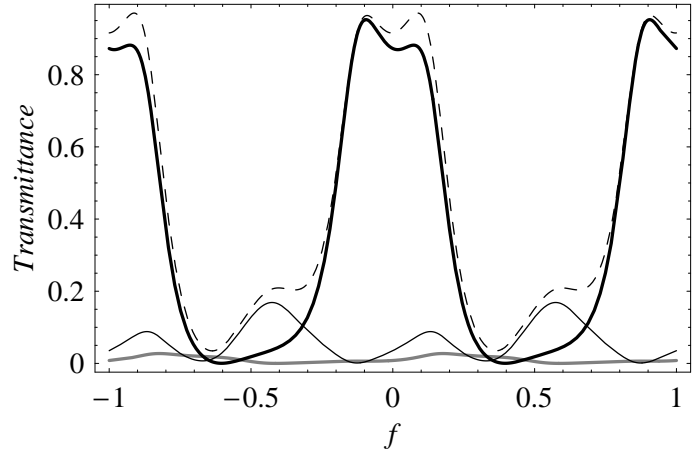}\\
\caption{Transmissions coefficients $\mathcal{T}_{11}$ (thick
black line), $\mathcal{T}_{21}$ (full line), and
$\mathcal{T}_{01}$ (thick grey line) as a function of the
Aharonov-Bohm flux $f$ with the following choice of parameters:
$kL=3.5$, $\gamma=0.2$, $pL=3$. The dashed curve represents the
total transmission including also 3 and 4 phonons contribution.}
\label{fig:coeff_vs_flux}
\end{figure}

\begin{figure}[htp]
\centering
\includegraphics[scale=.4]{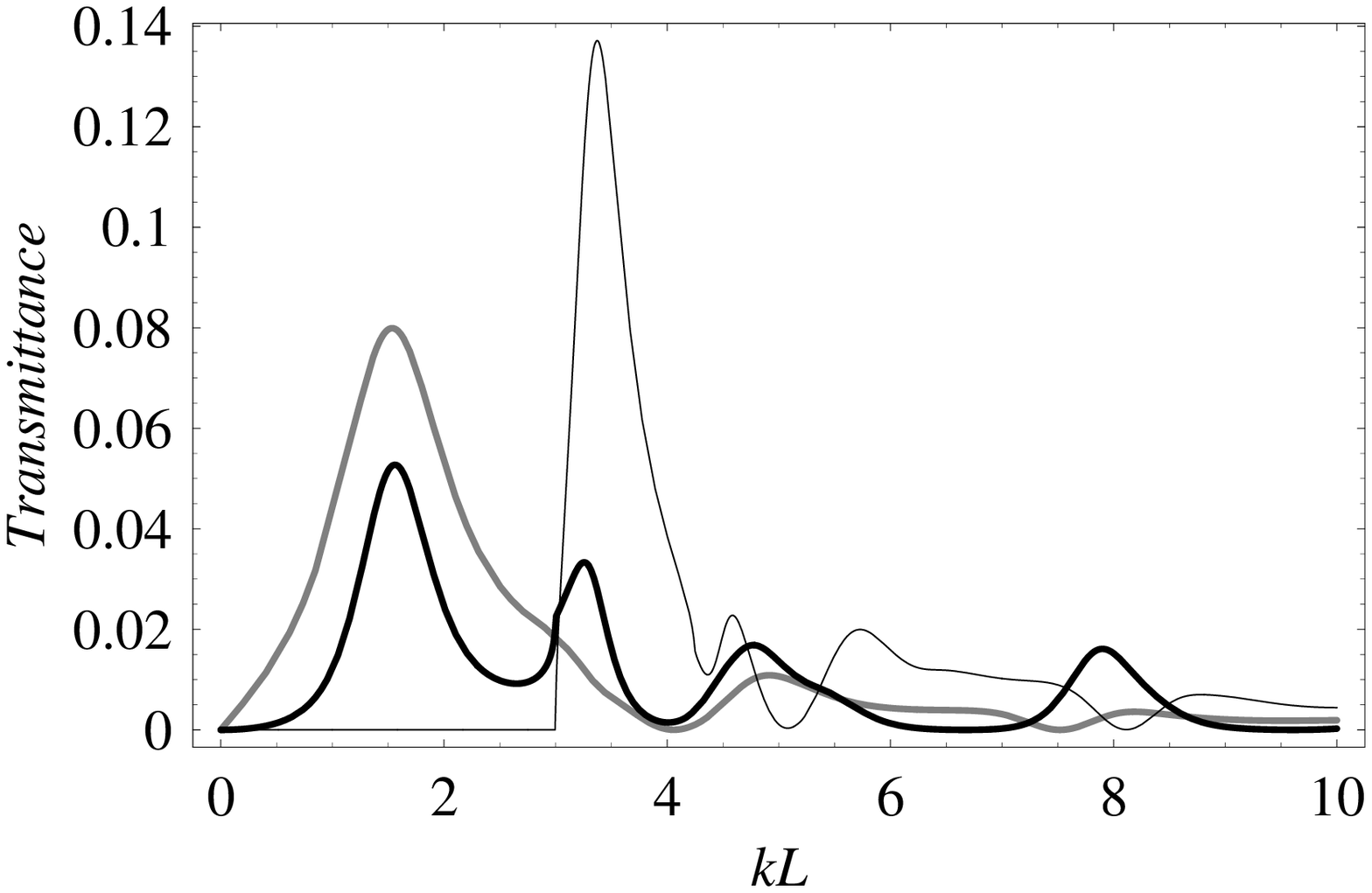}\\
\includegraphics[scale=.425]{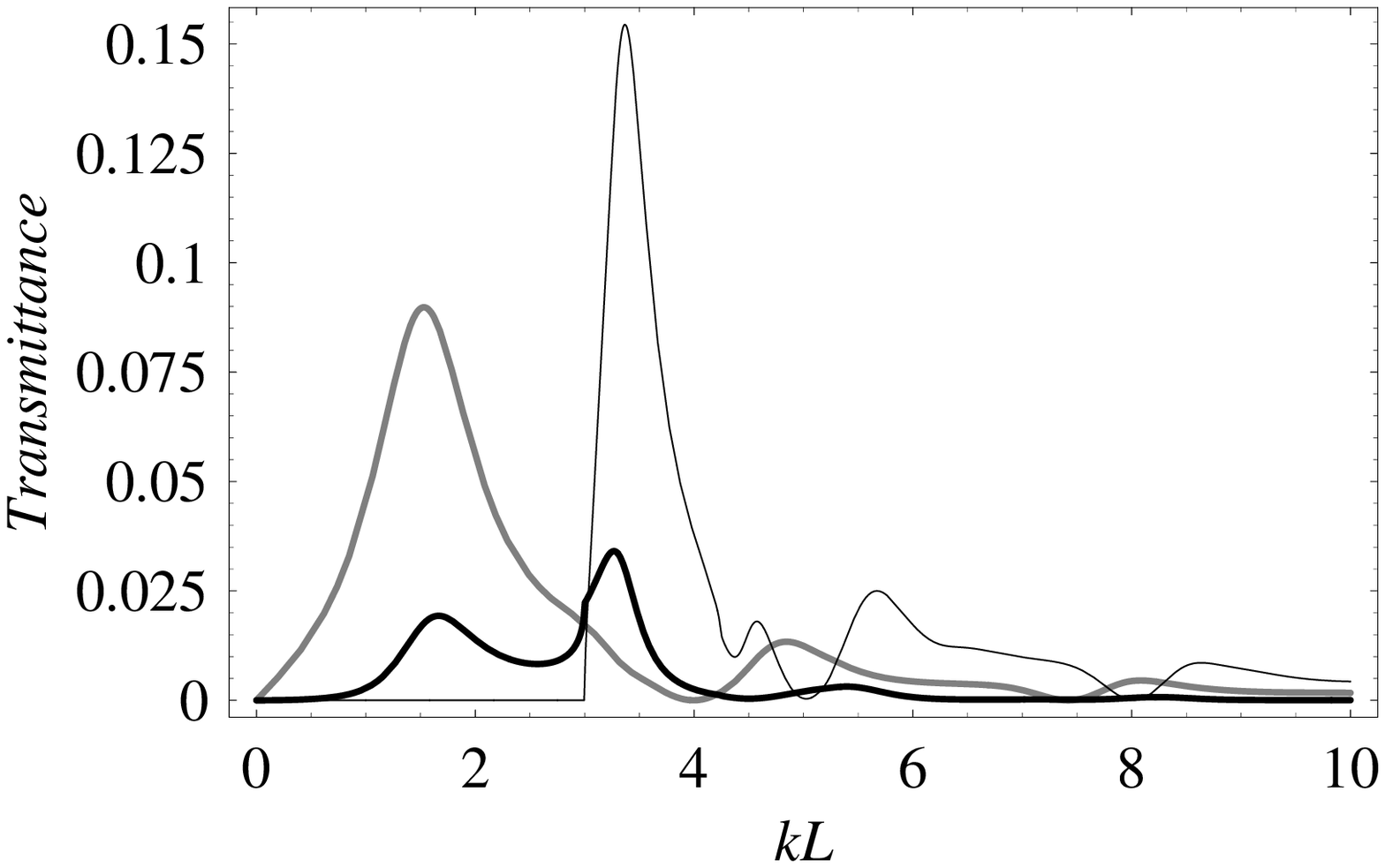}\\
\includegraphics[scale=.4]{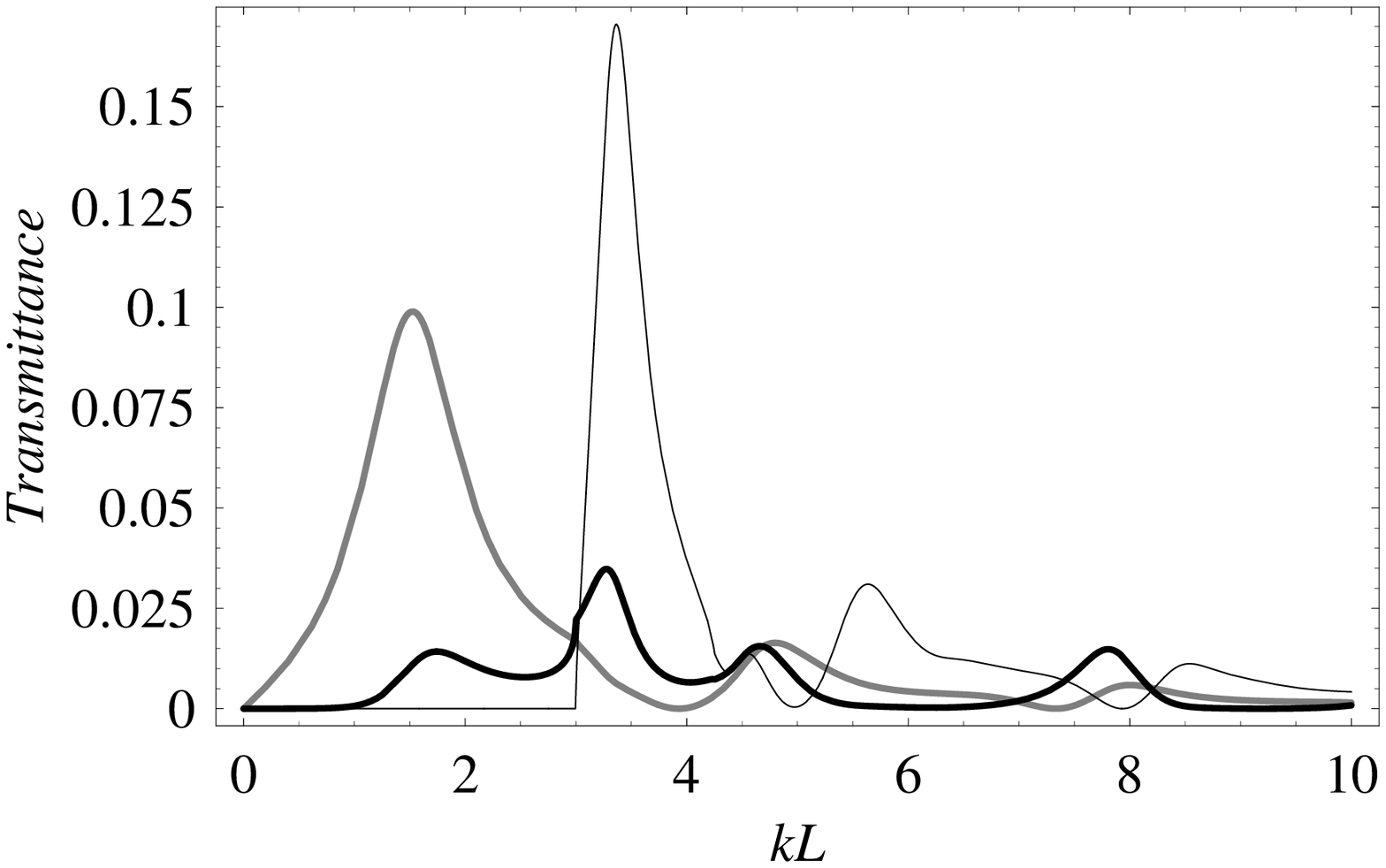} \hfill
\caption{Transmissions coefficients $\mathcal{T}_{11}$(thick black
line), $\mathcal{T}_{21}$ (full line), and $\mathcal{T}_{01}$
(thick grey line) as a function of $kL$ for the following choice
of parameters: $\gamma=0.2$, $pL=3$ and $f=0.49,0.5,0.51$(top to
bottom).}\label{fig:coeff_vs_en}
\end{figure}

In order to solve the linear system (\ref{eq:linear_syst}) for the
unknown transmission and reflection coefficients, a pruning
procedure has been applied (similar to Ref.\onlinecite{bonca99}).
The method is based on the simple idea of fixing the number of
phonon modes at $s=S$, and than iteratively removing all the
coefficients with phonon indices greater than $S$, adjusting the
remaining transmission and reflections coefficients in such a way
to have a total probability (reflection and transmission) equal to
one. Once fixed $s=S$, the problem reduces to solve a linear
system in the variables
{$r_{sl},t_{sl},a_{sl},b_{sl},A_{sl},B_{sl}$}, with
$s=0,\ldots,S$. In  the following we consider $S=4$. Once the
scattering problem is solved, the transmission probability, both
for elastic and non-elastic processes, is defined as follows:

\begin{equation}
\mathcal{T}_{ml}=|t_{ml}|^2\frac{k_m}{k_l},
\end{equation}
where $\mathcal{T}_{ml}$ represents the transmission probability
from the $l$-phonon state to the $m$-phonon state. In what follows
we assume that the starting phonon state is $l=1$, and the initial
electron momentum $k_l$ is fixed to $k$.  The total transmission
probability is obtained as $\sum_m \mathcal{T}_{m1}$.
We solved the scattering problem for $s=0,1,2,3,4$ phonons,
assuming the initial state of the cantilever corresponding to the
single phonon state ($l=1$). In Fig.\ref{fig:coeff_vs_flux}, the
transmission coefficients and the transmittance are shown as a
function of the external flux $f$ for $kL=3.5$, $\gamma=0.2$ and
$pL=3$. Even tough the main contribution to the transmission is
generally due to the elastic term $t_{11}$, a strong influence of
the inelastic term $t_{21}$ is observed close to half-integer
values of the flux. The lifting from zero of the conductance close
to an half-integer flux is the signature of a phonon-assisted
tunnelling (PAT). To better discern elastic from inelastic
contributions to the scattering in Fig.\ref{fig:coeff_vs_en} the
transmission coefficients $t_{11}, t_{21}, t_{01}$ are plotted as
a function of $kL$ for $\gamma=0.2$, $pL=3$ and $f=0.49, 0.5,
0.51$ (from top to bottom). As above, a strong competition between
elastic and inelastic scattering amplitudes is observed close to
half-integer values of the flux. A general feature of all the
panels in Fig.\ref{fig:coeff_vs_en} is that in the low energy
region ($kL<3$) the scattering amplitude is dominated by the
coefficient $t_{01}$ describing the process of emission of a
phonon. In the intermediate energy region ($3<kL<5$) the
transmission appears strongly affected by a resonant peak related
to absorption of a phonon ($t_{21}$), while in the high energy
region ($kL>5$) an alternating behavior is observed. Further, in
Fig.\ref{fig:coeff_vs_en}(top panel), close to $kL=4$, only the
inelastic coefficient $t_{21}$ contributes to the transmission.
This follows from the fact that an electron can be transmitted
only by changing the cantilever state, i.e. by means of the
absorption/emission of a phonon in the final state. Similar
behavior can be seen for other initial momenta $kL$.

\begin{figure}[htbp]
\centering
\includegraphics[scale=.55]{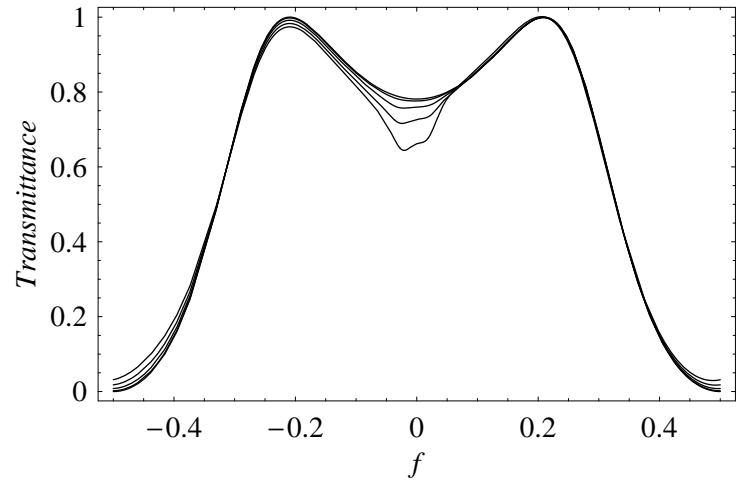}\\
\caption{Total transmission amplitude as a function of the total
flux $f$. The curves correspond to the following choice of
parameters: $kL=5.5$, $pL=3$ and $\gamma$ from $0$ to $0.2$ (top
to bottom with step $0.05$).} \label{fig:asym}
\end{figure}

\begin{figure}[htbp]
\centering
\includegraphics[scale=.369]{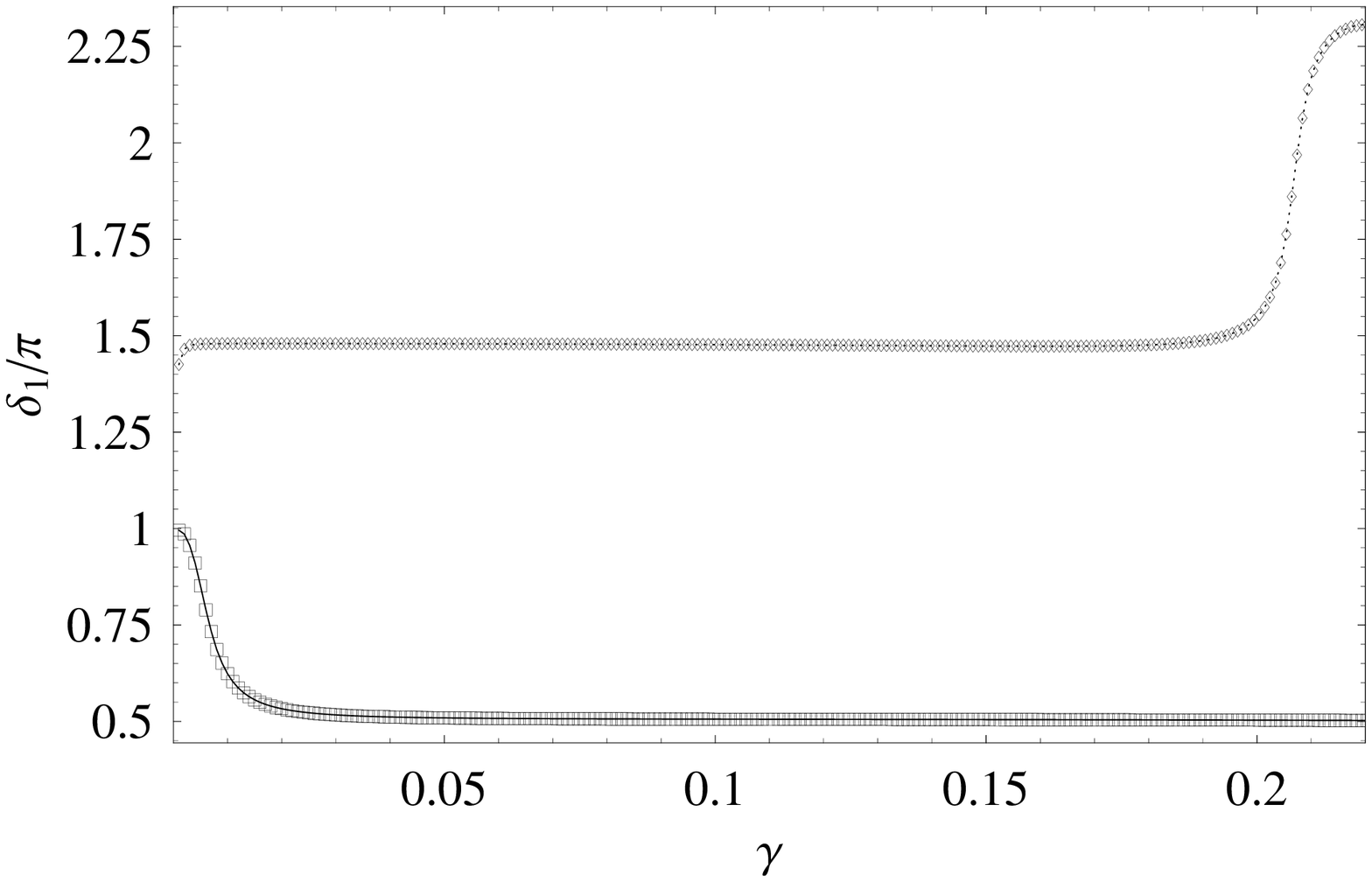}
\includegraphics[scale=.35]{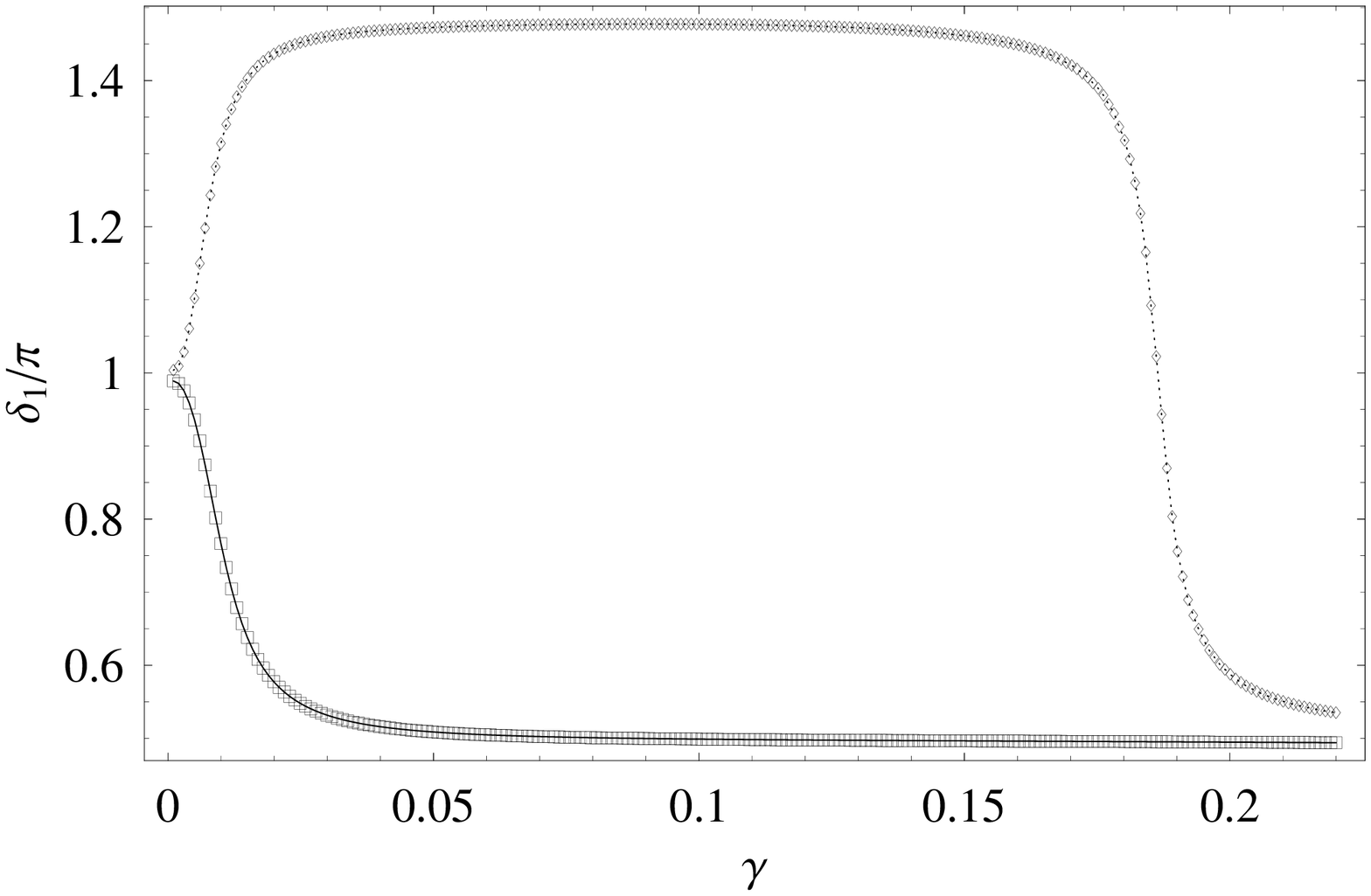}
\\
\caption{Phase shift $\delta_1$ computed for: $kL=7$ (box) and
$kL=5.5$ (diamond) [left panel] and $kL=8.9$ (box) and $kL=9.9$
(diamond) [right panel] by fixing $pL=3$. In all the curves
$\delta_1=\pi$ when $\gamma=0$.} \label{fig:phase_shifts}
\end{figure}

Let us apply the general conductance formula in the linear
response regime (Landauer-B\"{u}ttiker
formula\cite{landauer_cond}), to investigate the dependence of the
transmission on the flux $\phi$ and through that obtain the
behavior of the phase shift. The transmission phase through a
mesoscopic AB ring has the general property of phase rigidity
coming from the two-terminal nature of the set-up and which is
generally based on time-reversal symmetry-breaking and current
conservation\cite{buttiker_prl_86}, i.e. the transmission
probability amplitude satisfies the property $\mathcal{T}_{\alpha
\beta}(kL,\phi)=\mathcal{T}_{\beta \alpha}(kL,-\phi)$ as a
function of the flux and of the incident electron energy
($\alpha,\beta=L,R$ being the lead index). Combining time-reversal
symmetry requirement with current conservation
$|\mathcal{T}_{\alpha \beta}(kL,\phi)|^2=|\mathcal{T}_{\beta
\alpha}(kL,\phi)|^2$, implies that the linear conductance
$\mathcal{G}$ is an even function of the flux, whose Fourier
transform is
\begin{equation}
\label{eq:ft} \mathcal{G}(\phi)=\mathcal{G}_0+\sum_n \mathcal{G}_n
\cos(n \phi+\delta_n),
\end{equation}
and $\phi=2 \pi f$.  Obviously, the phase shift $\delta_n$ can
take only two values $0$ and $\pi$\cite{note_hilbert_transf}.
Thus, the phase of the two terminal AB ring has to be rigid, or
change abruptly by $\pi$ as the accumulated phase in one arm is
being varied. This peculiar behavior, known as phase rigidity has
been studied
extensively\cite{imry_review,buks_exp,yeyati_AB,phase_rig_1,phase_rig_2,phase_rig_3,sun99,leturcq06exp}.
Experimentally\cite{buks_exp,exp_kobayashi1,exp_kobayashi2} and
theoretically\cite{jayannavar} various methods of avoiding
phase-rigidity in rings and ring-dots systems were previously
discussed. Here we are proposing, for the first time, observation
of phase rigidity breaking in a nanoelectromechanical system as an
alternative way of observing the continuous variation of the
transmission phase through a two-terminal mesoscopic system. Our
results in Fig.\ref{fig:asym} show that for different values of
the interaction $\gamma$ inclusion of inelastic scattering, breaks
phase rigidity\cite{note_onsager,kang_onsager}. The first evidence
for this comes from the fact that $\mathcal{T}$ is non symmetric.
This is clearly seen in the vicinity of the antiresonances in the
conductance, which are approximately located at the energy values
determined by the electron-phonon coupling $\gamma$:
\begin{eqnarray}
\label{eq:antiresonaces}
K^l_{nm}L \approx \sqrt{\pi^2(n+f)^2+(p
L)^2(l-m+\gamma^2)},
\end{eqnarray}
where $n$ labels the momentum of the eigenstates in the ring,
$m(l)$ labels the final(initial) phonon channel. The amount of
phase rigidity violation depends on strength of $\gamma$ and it is
related to the momentum of incoming electron. For instance, in
Fig.\ref{fig:asym}, the total transmission as a function of the
flux $f$ is shown, for $pL=3$, $kL=5.5$ and varying $\gamma$ from
zero to $0.2$ (from top to bottom with step 0.05).  Since the
phase shift is a measurable quantity in the experiments, in
Figs.\ref{fig:phase_shifts}, the behavior of $\delta_1$ is shown
as a function of $\gamma$ for different values of the
antiresonance momenta. As shown, in the absence of interaction,
i.e. under Onsager symmetry, the phase shift is equal to $\pi$,
corresponding to a local minimum in the conductance versus flux
close to $f=0$. When the interaction is turned on, the phase shift
$\delta_1$ converges quite rapidly to the values $\pi/2$ or
$3\pi/2$, depending on the value of the momentum $kL$. The
interaction region characterized by a rapid variation of the phase
shift can be exploited in experiments to obtain information on the
electron-phonon coupling simply by measuring the phase shift of
the linear conductance. Furthermore, since this analysis can be
done for all the antiresonances, the value of the interaction
$\gamma$ can be obtained by experimental measurements of the phase
shift. The phase rigidity breaking we have discussed, has close
similarities to that described by Sun \textit{et al.} in
[\onlinecite{sun99}] where a two-terminal modified AB ring with a
quantum dot inserted in one arm has been considered. There, the
phase rigidity is broken in the linear regime by applying a
time-varying microwave (MW) field on the quantum dot. Further, the
behavior of the phase shift in Figs.\ref{fig:phase_shifts} is
similar to that shown in Fig. 4 of Ref.\onlinecite{leturcq06exp}
for non-linear transport in an AB ring. Even tough, the
experimental analysis as a function of a perpendicular magnetic
field is focused on nonlinear contribution to the conductance, the
analogy in the phase shift behavior can be associated to an
effective energy shift in the lower arm provided in our case by
the interaction with the cantilever, while there is due to a gate
voltage.
%

In conclusion, we have analyzed the transmittance and the phase
shifts behavior in a two-terminal AB ring whose lower arm is
coupled to a cantilever. We showed that the phase rigidity can be
broken as an effect of the inelastic scattering channels
introduced by the cantilever, and a continuous phase shift can be
obtained through the measurements of the linear conductance. The
continuous phase variation as a function of the incident electron
energy in experiments can be exploited to obtain the value of the
electron-phonon coupling. Our proposal is within reach with
today's technology employed in nanoelectromechanical systems. It
can be realized by means of a semiconducting ring (e. g. In As
\cite{hui}) operating at mK temperatures with radius $R \approx 60
\div 100 nm$ coupled to a molecular cantilever (e. g. a
cantilevered or bridged single walled carbon nanotube
\cite{chunyu}) with a mass $\approx 10^{-24} \div 10^{-23} kg$ and
with fundamental frequency $\omega \approx 10^{2} \div 10^{3} GHz
$, thus allowing to realize dwell times for the electrons less or
of the order of the cantilever frequency.
The cantilever length can be taken of the order of the
(0.02-0.05)$\mu$m and its distance from the arm of the ring of the
order (0.01-0.1)$\mu$m. In this way an extra electron in
correspondence of the position of the tip causes a small vertical
displacement. The phase rigidity breaking can be regarded a common
feature of two-terminal nanomechanical systems and thus we propose
measurements of phase shifts as a way to determine the strength of
the electron-phonon coupling in NEMS.


\section{Acknowledgments}
We acknowledge Fabio Pistolesi for his enlightening comments and
careful reading of the paper. F. Romeo thanks the laboratory LPMMC
of Grenoble for kind hospitality.




\begin{thebibliography}{99}
\bibitem{note_hilbert_transf} The scattering phase-shift can
also be determined by the Hilbert transform of the scattering
probability as shown in Ref.[14].
\end{thebibliography}
\end{document}